\documentclass{imamat}

\usepackage{mwe}
\usepackage{amsmath} 
\usepackage{amsthm} 
\usepackage{cite} 
\usepackage{graphicx} 
\usepackage{subfig}  
\usepackage{xcolor}

\begin{document}

\title{
Snaking without subcriticality: grain boundaries as non-topological defects}

\shorttitle{Snaking of defects} 
\shortauthorlist{Subramanian, Archer, Knobloch, Rucklidge} 

\author{
\name{Priya Subramanian$^*$}
\address{Mathematical Institute, University of Oxford, Oxford OX2 6GG, United Kingdom \email{$^*$Corresponding author: priya.subramanian@maths.ox.ac.uk}}
\name{Andrew J. Archer}
\address{Department of Mathematical Sciences and Interdisciplinary Centre for Mathematical Modelling, Loughborough University, Loughborough LE11 3TU, United Kingdom}
\name{Edgar Knobloch}
\address{Department of Physics, University of California at Berkeley, Berkeley, California 94720, USA}
\and
\name{Alastair M. Rucklidge}
\address{Department of Applied Mathematics, University of Leeds, Leeds LS2 9JT, United Kingdom}
}

\maketitle

\begin{abstract}
{Non-topological defects such as grain boundaries abound in pattern forming systems, arising from local variations of pattern properties such as amplitude, wavelength, orientation, etc. We introduce the idea of treating such non-topological defects as spatially localised structures that are embedded in a background pattern, instead of treating them in an amplitude-phase decomposition. Using the two-dimensional quadratic-cubic Swift--Hohenberg equation as an example we obtain fully nonlinear equilibria that contain grain boundaries which are closed curves containing multiple penta-hepta defects separating regions of hexagons with different orientations. These states arise from local orientation mismatch between two stable hexagon patterns, one of which forms the localised grain and the other its background, and do not require a subcritical bifurcation connecting them. Multiple robust isolas that span a wide range of parameters are obtained even in the absence of a unique Maxwell point, underlining the importance of retaining pinning when analysing patterns with defects, an effect omitted from the amplitude-phase description.
}
{non-topological defect, grain boundaries, penta-hepta defect, Swift--Hohenberg equation, Spatial localisation}
\\
2000 Math Subject Classification: 35B36, 74N20, 70K50
%
\end{abstract}

\section{Introduction and Motivation}


Defects are of fundamental importance in the study of patterns as well as in materials science \citep{mermin1979topological, ch93, chaikin1995principles} and may take the form of domain walls, grain boundaries, dislocations and disclinations. These structures can be stationary or glide or climb or otherwise move through the pattern, and are present in one, two and three dimensions.  One traditional approach to the study of defects is using an amplitude-phase decomposition in which a periodic pattern is described by a homogeneous amplitude and defects are associated with zeros of the amplitude. Since the phase at these locations is undefined these defects are known as topological defects and these have attracted the greatest attention \citep{mermin1979topological}. Such defects are characterised by a non-zero topological charge, proportional to the integral of the phase around the defect. However, many defects are non-topological in character, but are nonetheless of fundamental significance \citep{chaikin1995principles}. For these defects the phase is defined everywhere and the topological charge vanishes. Non-topological defects are frequently believed to be less important since isolated defects of this type may heal or otherwise disappear without interaction with another defect. However, this is not always the case, and if the defect has internal structure it may persist as a result of frustration or locking to the background state.

In the present work we adopt a different point of view and think of defects as spatially localised structures embedded in a background pattern. In general, defects represent a local disruption of the background state, and so can consist of holes where its amplitude is locally reduced, or local changes in the background wavenumber or its orientation. In most cases the background state is a periodic or crystalline state, but defects in quasicrystals or even disordered media also arise \citep{korkodi2013}.

\begin{figure}
\centering{\includegraphics[height=0.4\linewidth]{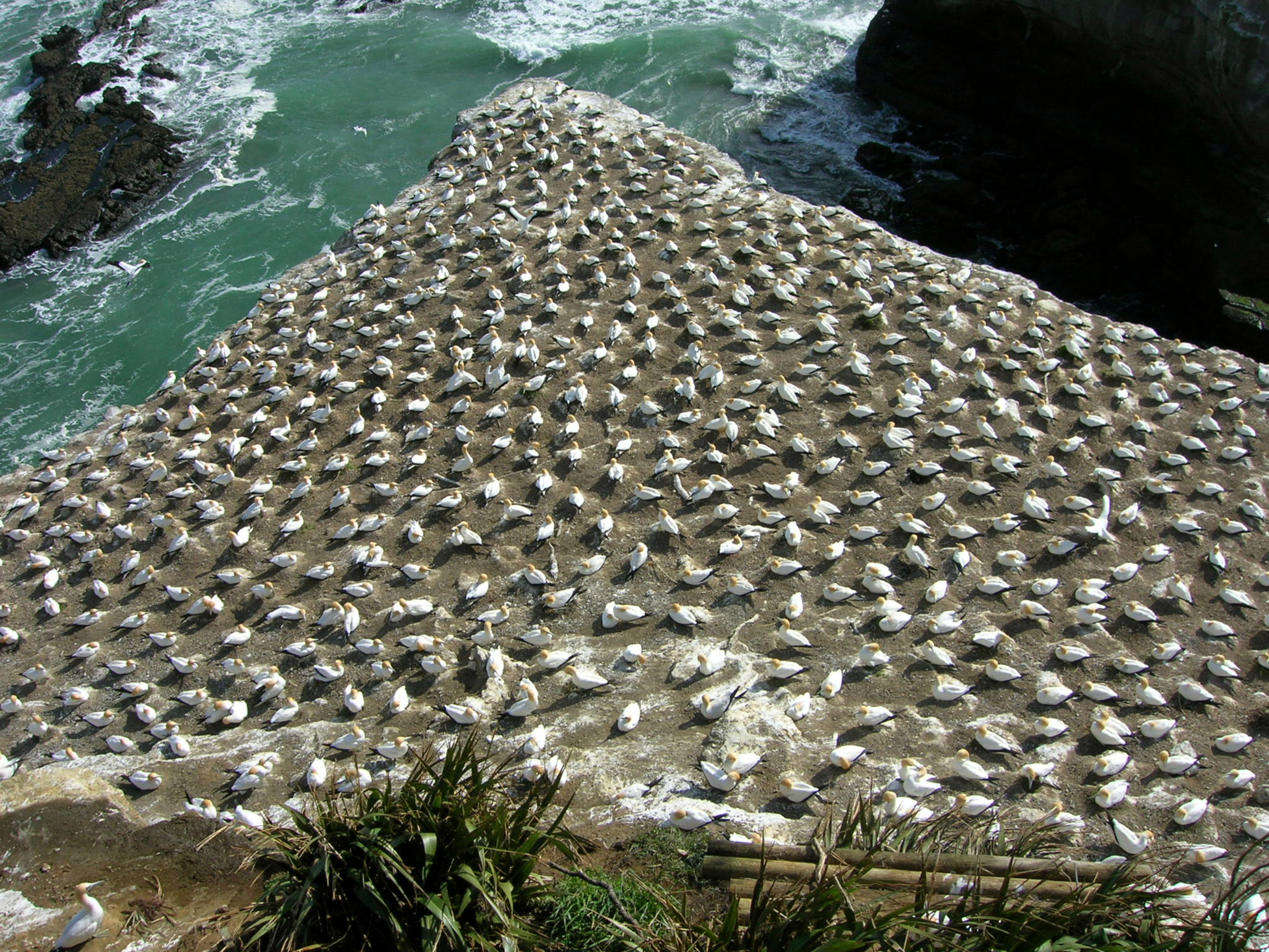}
\hspace{0.3cm}
\includegraphics[height=0.4\linewidth]{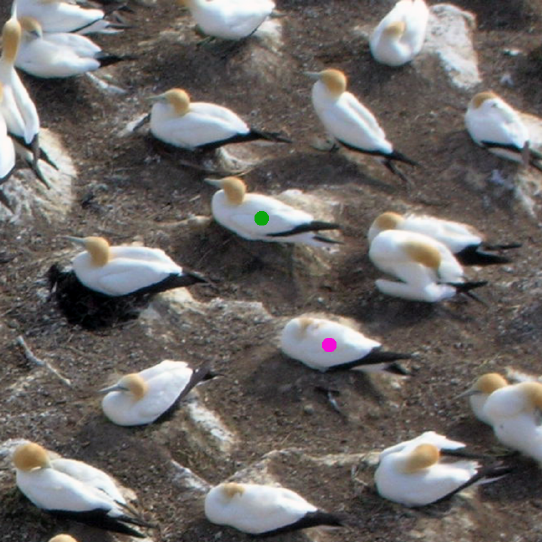} } 
\caption{When neighbours keep their distance: gannets nesting at Muriwai beach, New Zealand. The image on the left shows the overall hexagonal ordering of the birds while that on the right shows a penta-hepta defect (PHD). The green (pink) markers identify gannets with five (seven) neighbours instead of the usual six. Photo credit: Barbora Knobloch.
}
\label {fig:0}
\end{figure}

Spatial localisation has been the subject of numerous recent studies (see \citet{knobloch2015spatial} for a review) leading to the following picture of such structures. Defects may be described as holes in a pattern and these bifurcate from a pattern branch near its folds and are initially unstable. Both of these facts are a consequence of the presence of a Floquet multiplier $+1$ at such folds. In variational systems the resulting hole branch undergoes homoclinic snaking \citep{woods1999heteroclinic} in the vicinity of a Maxwell point, the parameter value at which the free energy associated with the pattern is the same as that of a competing state. Homoclinic snaking is responsible for stabilising some of the hole states and is a consequence of the locking of a front between the two competing states to the heterogeneity or spatial structure on one or both sides of the front. On the basis of this picture we expect stable defect states to be present in regions of bistability, and associated with the presence of a Maxwell point, i.e., in the so-called snaking or pinning region. In particular, we expect holes to come in the form of infinite families of defect structures of specific symmetry and ever-increasing size, all of which coexist stably within the pinning region. In general, these structures either snake or lie on a stack of isolas \citep{beck09}. Much of the above picture has been developed on the basis of detailed studies of the bistable Swift--Hohenberg equation \citep{burke2006localized}, a prototypical pattern-forming equation exhibiting bistability between a pattern state and a spatially homogeneous state. 

We point out that the traditional approach to analyse defects using an amplitude-phase decomposition ignores the possibility of pinning thereby collapsing the family of defects to a stationary heteroclinic front at the Maxwell point. The present work shows that pinning must be retained in studies of defects and that its effects in general extend over a broad range of parameter values. To demonstrate the importance of pinning and the associated multiplicity of defect structures we adopt the two-dimensional quadratic-cubic Swift--Hohenberg equation (SH23) and focus on one of the most important defect structures in two dimensions, the penta-hepta defect (PHD) in a hexagonal pattern. This classical structure consists of a bound state of a peak with five neighbours and a peak with seven neighbours, instead of the usual six neighbours in a defect-free hexagonal pattern. Figure \ref{fig:0} shows such a defect in a largely hexagonal arrangement of gannet nests at Muriwai beach in New Zealand.

In particular, we compute here three distinct defect structures and the associated isolas, thereby providing, compelling evidence for the coexistence of multiple defect structures in two dimensions. Earlier work in one dimension examined domain boundaries between two distinct spatially periodic states in SH357, the cubic-quintic-septic Swift--Hohenberg equation \citep{uecker19}, and confirmed the presence of defect snaking in one spatial dimension. Snaking of grain boundaries between stripe patterns with different orientations in 2D has been investigated in the Swift--Hohenberg equation with cubic nonlinearity \citep{LloydScheel2017}. In our work, we investigate the Swift--Hohenberg equation with quadratic and cubic nonlinearities in the supercritical regime where hexagonal patterns are stable. The PHDs we study are the result of a local change in the orientation of the hexagonal pattern. Thus, these defects are the result of bistability between a hexagonal pattern and a rotated version of the same pattern. We expect similar structures in mass-conserving systems described by the conserved Swift--Hohenberg equation, or equivalently the phase-field crystal (PFC) model, since stationary states of this model satisfy the same stationary equation as that studied below \citep{TARG2013pre}.

The bistable two-dimensional SH23 equation describes competition between hexagonal patterns and a trivial, spatially homogeneous state. The equation is variational with a Maxwell point between the hexagonal state and the trivial state. For this reason this equation admits both holes in a hexagonal pattern and hexagon patches embedded in the trivial background \citep{LloydSIADS2008}. Here we are interested in defects that are not associated with the trivial state and that are found in the supercritical regime. These defects are associated with grain boundaries between two hexagonal patterns, obtained by rotating the hexagonal pattern in part of the domain through 30 degrees relative to the rest. This procedure generates closed rings of PHDs. These are found to lie on distinct isolas despite the absence of a discrete Maxwell point between the two competing hexagonal states (both of which have the same free energy). We mention that an isolated PHD in a hexagonal pattern can be described using an amplitude-phase description \citep{tsimring95}. This description shows that such defects can be thought of as a bound state of two dislocations, of opposite signs, in two of the three Fourier modes involved in the basic hexagonal structure. Such defects can be stationary or can climb \citep{tsimring95}. However, the description of such motion is expected to be strongly influenced by the presence of pinning between of the PHD and the surrounding hexagonal pattern. It is this effect that is responsible for the multiplicity of PHD structures in this system.

As already mentioned, equilibria of the Swift--Hohenberg equation are also equilibria of the PFC model, which is a simple theory for the crystallisation of matter. In this context, the results presented below apply to the defects observed at grain boundaries in polycrystalline materials. In particular, since we are considering a two-dimensional system, our results apply to the grain boundaries in two-dimensional solids such as graphene \citep{hirvonen2016multiscale, hirvonen2017energetics}, in which the defects are of great importance near to melting \citep{zakharchenko2011melting} and indeed affect the very nature of melting in two dimensions \citep{halperin1978theory}.

This paper is organised as follows. In the next section we introduce the model equation we study, and the demodulation technique we use to isolate the defect state. In section 3 we describe our procedure for generating rings of PHDs. These structures are shown in section 4 and found to lie on distinct isolas. The spatial extent of the influence of the resulting PHD structures on the background hexagonal pattern is quantified. The paper concludes with a brief summary and some questions concerning the implications of multiple equilibria with defects for both pattern formation and material science that will, we hope, inform future work.


\section{Model and Methodology}

We consider the pattern-forming system SH23, a partial differential equation for a scalar field $u({\bf x},t)$,
\begin{equation}
\frac{\partial u}{\partial t} = \mu u - ( 1 + \nabla^2)^2 u +Q u^2 -u^3\,.
\label{eq:sh23}
\end{equation}
For some background on this equation, see e.g.\ \citet{burke2006localized}. The parameters in this model are the linear growth rate $\mu$ and the strength of the second order nonlinear interactions $Q$. The domain chosen in all the calculations in this work, unless specified otherwise, is a square with side length $L_x=L_y=60\pi$ (30 wavelengths), discretised with $256$ grid points in each direction. Time-stepping from a given initial condition is accomplished pseudospectrally using second-order exponential time differencing (ETD2) \citep{CoxJCP2002}. Numerical continuation is done using a pseudo arc-length continuation similar to that employed in \citet{SubramanianNJP2018}. Note that Eq.~\eqref{eq:sh23} describes non-conserved gradient dynamics ${\partial u}/{\partial t}=-{\delta \mathcal{F}[u]}/{\delta u}$, with the free energy functional
\begin{equation}
\mathcal{F}[u] = \int \left[ \frac{1}{2}\mu u^2 + \frac{1}{2} u( 1 + \nabla^2)^2 u -\frac{1}{3}Q u^3 +\frac{1}{4}u^4\right] \mathrm{d}\mathbf{x}\,.
\label{eq:free_energy}
\end{equation}

\begin{figure}
\centering{\includegraphics[width=0.95\linewidth]{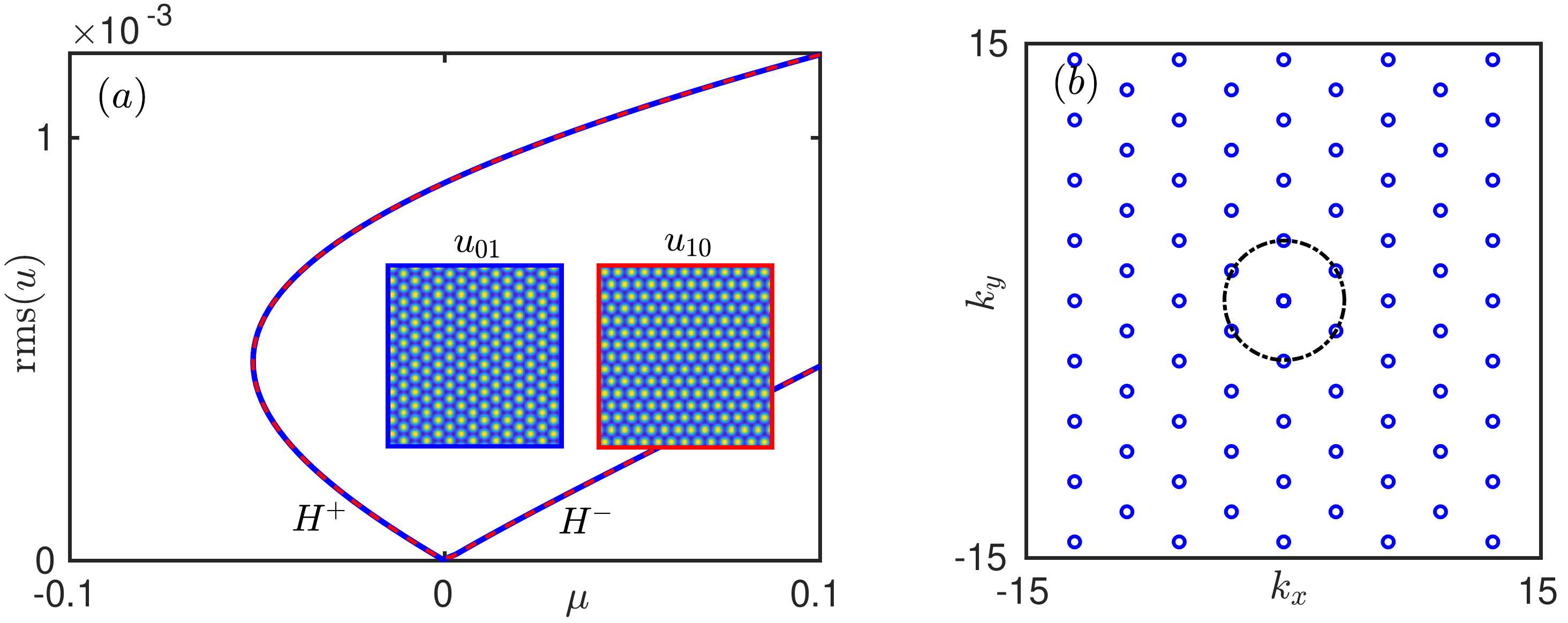} } 
\caption{($a$) Bifurcation diagram of extended hexagonal states showing the root mean square norm ${\rm rms}(u)$ computed for SH23 with $Q=0.75$. Insets show the two different orientations of the extended hexagonal states of interest, the state $u_{01}$ with a blue border and $u_{10}$ with a red border; both have the same norm and correspond to solutions from the $H^{+}$ branch at $\mu=0.25$. ($b$) Components of the reciprocal lattice vectors (RLVs) of the state $u_{01}$ which are set to zero in the filter $\mathcal{P}$, shown in a periodic domain of size 30 wavelengths (256 grid points) in each direction. The dashed-dotted black circle passes through the first order RLVs. 
}
\label {fig:1}
\end{figure}

Figure \ref{fig:1}($a$) shows the bifurcation diagram of extended hexagonal states H$^\pm$ (so-called up-hexagons and down-hexagons, respectively) as a function of the chosen bifurcation parameter $\mu$ when $Q=0.75$. A transcritical bifurcation occurs at $\mu=0$, generating H$^+$ for $\mu<0$ and H$^-$ for $\mu>0$. The H$^+$ states undergo a saddle-node bifurcation at $\mu=-0.05$. Shown are two superposed branches differing in the orientation of their wavevectors in their Fourier spectrum. The blue line has solutions that include the wavevector $(0,1)$, whose real space pattern is shown in the inset with a blue border and labeled $u_{01}$. Solutions along the dashed red line have wavevectors including $(1,0)$ whose real space pattern is shown in the inset with a red border and labeled $u_{10}$. We observe that the root mean square values of both these extended states are equal, as expected since the patterns are just rotations of each other.

We can differentiate between the two extended hexagonal patterns if we consider one of them to be the reference state, say $u_{01}$, and measure the norm of the difference between a given state and $u_{01}$. In order to identify the parts of a given field $u$ that differ from a chosen reference state, we create a filter matrix $\mathcal{P}$ in spectral space from the reference state $u_{01}$ in the following way. 

First, we identify all the Fourier components that contribute nontrivially to the spectrum of the reference state by identifying the reciprocal lattice vectors (RLVs) of the reference state $\hat{u}_{01}$, including wavevectors that are obtained via higher order interactions (see e.g.\ the discussion in chapter 4 of \citet{chaikin1995principles}). Figure \ref{fig:1}($b$) shows the locations of these wavevectors in the 2D domain as blue circles. The dashed-dotted circle has been added to help identify the order 1 RLVs of magnitude $|\bm{k}|=1$, the radius of the circle. In order to calculate the distance of a given state from this hexagonal pattern, we seek to set all contributions from the above Fourier components to be zero in the pattern of interest. In this way we define the projection $\mathcal{P}$ whose application to the Fourier transform of a given field $u({\bf x})$ sets all the Fourier components identified in Fig.~\ref{fig:1}($b$) to zero. The resulting field $\textrm{IFFT}(\mathcal{P}\hat{u})$, where IFFT denotes the inverse fast Fourier transform, is now spectrally filtered and has no components along the Fourier components of the reference field, in this case $u_{01}$.

\begin{figure}
\centering{\includegraphics[width=0.6\linewidth]{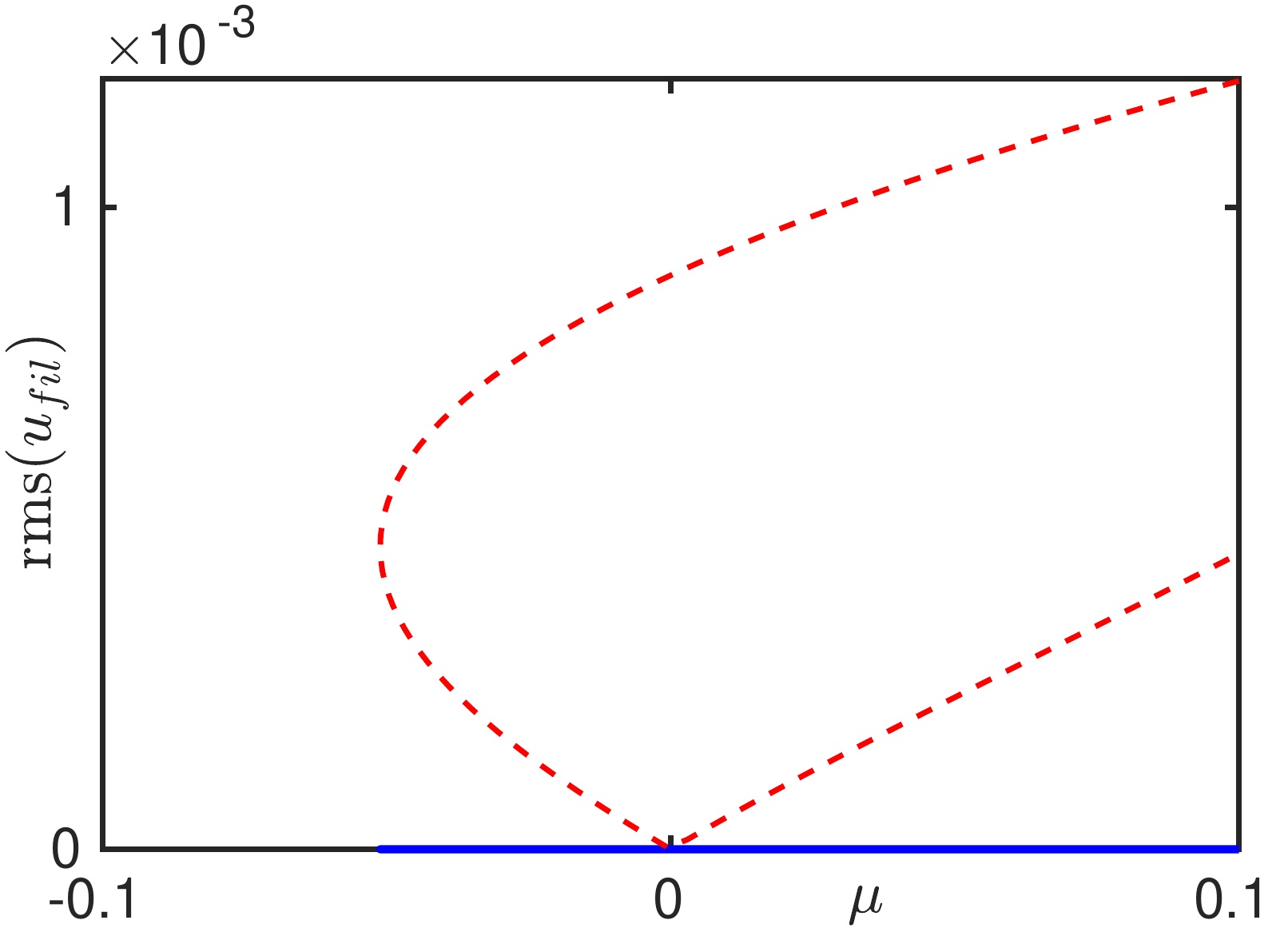} } 
\caption{Bifurcation diagram of extended hexagonal states showing the filtered norm ${\rm rms}(u_{fil})$ for $u_{10}$ (red) and $u_{01}$ (blue) as a function of the bifurcation parameter $\mu$. Other parameters are the same as in Fig.\,\ref{fig:1}.
}
\label {fig:2}
\end{figure}

If we consider the same extended hexagonal patterns as in Fig.~\ref{fig:1}($a$) and replot them using the root mean square of the spectrally filtered state $u_{fil}$, calculated as
\begin{equation}
{\rm rms}(u_{fil}) = \sqrt{\frac{1}{A} \int \left[\textrm{IFFT}( \mathcal{P}\hat{u}) \right]^2 \,dA}\,,
\label{eqn:specfilter}
\end{equation}
we obtain Fig.~\ref{fig:2}. Since the new measure ${\rm rms}(u_{fil})$ represents the departure of a pattern $u$ from another chosen reference pattern, in this case $u_{01}$, the state $u_{01}$ corresponds to the blue horizontal line with zero amplitude, while the state $u_{10}$, shown in red, remains unmodified at the scale of the figure (compare the red dashed lines in Figs.\,\ref{fig:1}($a$) and \ref{fig:2}). This is because in the case of extended periodic patterns, the higher order contributions to the Fourier spectrum of the pattern $u_{10}$ at wavevectors that are part of the higher order spectrum of the pattern $u_{01}$ are very small. The largest change arises from the removal of the bulk mode at $|\bm{k}|=0$. We mention that if the reference state is chosen to be the flat state, the measure (\ref{eqn:specfilter}) reduces to the usual ${\rm rms}(u)$. 

In this section we have introduced the idea of spectral filtering to help visualize states with defects. In the rest of this work we focus on the formation of spatially localised patterns that involve penta-hepta defects separating the two hexagonal states $u_{01}$ and $u_{10}$. However, the idea of spectral filtering to visualize defects is more general and can be used to describe more complex grain boundaries. 


\section{Penta-hepta defects and grain boundaries}

\begin{figure}
\centering{\includegraphics[width=0.95\linewidth]{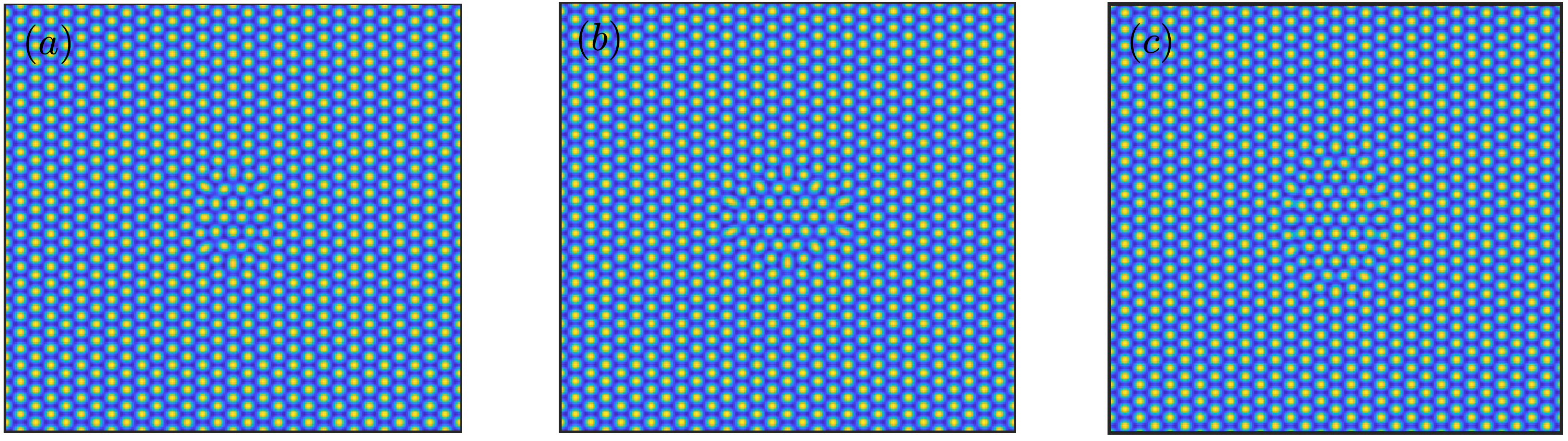} }
\caption{Coexisting equilibria with penta-hepta defects separating regions of hexagons with different orientations in the SH23 system with parameters $\mu=0.25$ and $Q=0.75$. All are dynamically stable states.
}
\label {fig:3}
\end{figure}

In the SH23 equation introduced in the previous section, we look for equilibria that involve penta-hepta defects. In order to promote the formation of hexagonal patterns, we set $Q=0.75$ and choose a positive value of the growth rate, $\mu=0.25$. At these parameters, we use a combination of time stepping and numerical continuation to compute multiple coexisting states that involve spatial localisation of a patch of $u_{10}$ within a $u_{01}$ background. Figure \ref{fig:3} shows several such examples that are dynamically stable as confirmed by time-evolving small perturbations using Eq.~\eqref{eq:sh23}.

\begin{figure}
\centering{\includegraphics[width=0.95\linewidth]{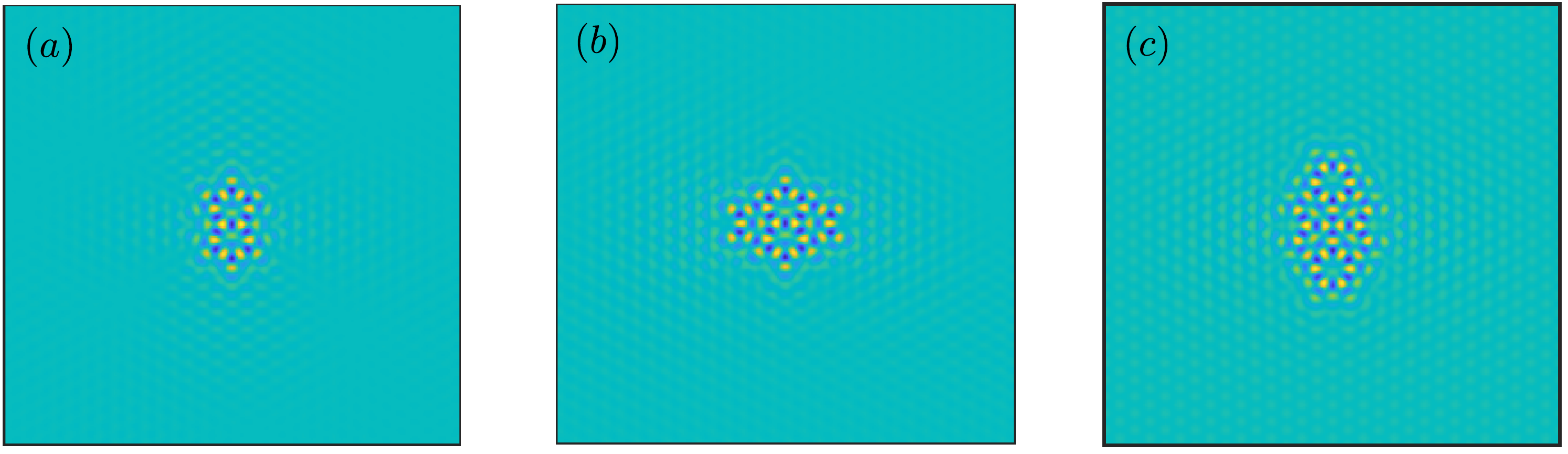} }
\caption{Corresponding spectrally filtered fields of the patterns in Fig.\,\ref{fig:3} showing locations in the $(x,y)$ plane with deviation from the background state $u_{01}$.}
\label {fig:4}
\end{figure}

Since it is difficult to identify visually the location of the defects in this state, we also look at the corresponding spectrally filtered fields $u_{fil}$ in Fig.\,\ref{fig:4}. Here we see that the regions in Fig.\,\ref{fig:3} that are purely hexagonal with orientation $u_{01}$ are transformed into flat (green) regions in the spectrally filtered plot. Additionally, peaks in the $u_{10}$ patch coinciding with those of $u_{01}$ are also removed, see e.g. the flat/green regions in the middle of the hexagonal ring of peaks in Fig.\,\ref{fig:4}($a$). Note also that as we move across the three panels, the size of the $u_{10}$ patch in the middle occupies a larger part of the domain. However, all of the cases shown here possess symmetry with respect to reflections in the $x$- and $y$-axes. 

\begin{figure}
\centering{\includegraphics[width=0.95\linewidth]{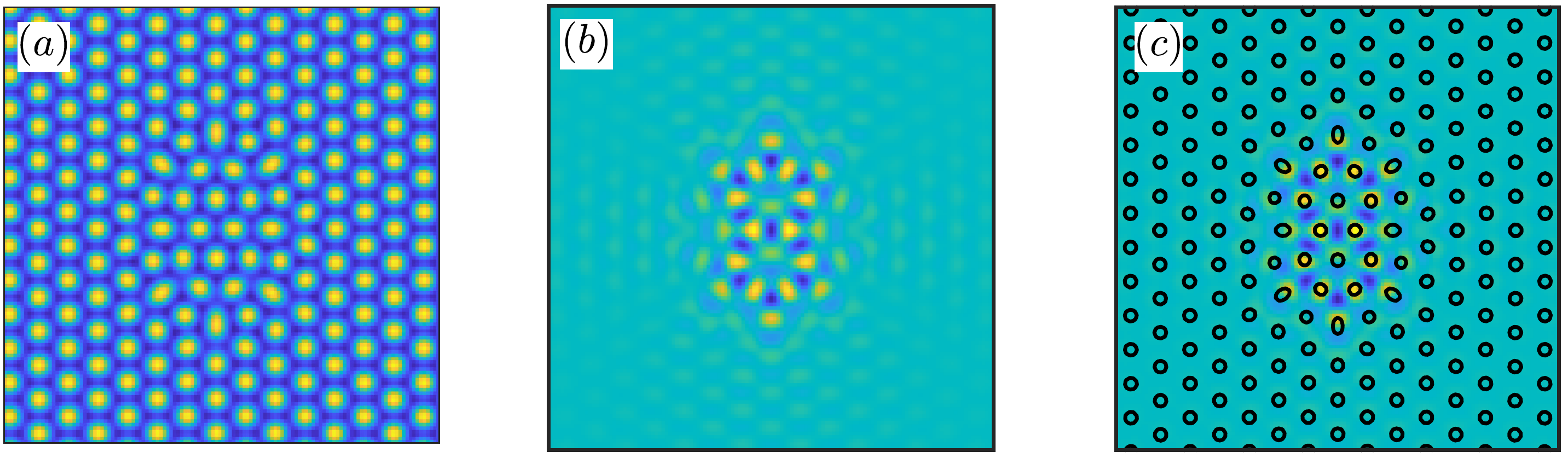} }
\caption{($a$) Zoomed-in view of the centre of the domain from Fig.\,\ref{fig:2}($a$) showing $u(x,y)$. ($b$) Corresponding zoomed-in area from Fig.\,\ref{fig:3}($a$) showing $u_{fil}$. ($c$) Contours determined at $65\%$ of $\max(u)$ from panel ($a$) shown in black, superposed on $u_{fil}$ from panel ($b$).
}
\label {fig:5}
\end{figure}

In order to correlate the location of the grain boundaries and associated penta-hepta defects in the $u$ fields shown in Fig.\,\ref{fig:3} with the corresponding peaks in Fig.\,\ref{fig:4}, we compare the two fields in panels ($a$): we zoom in to the centre of the domain in Fig.\,\ref{fig:3}($a$) and show the results in Fig.\,\ref{fig:5}($a$). Figure \ref{fig:5}($b$) shows the corresponding zoomed-in view of Fig.\,\ref{fig:4}($a$). In order to combine the information in both these figures, we estimate contours from panel ($a$) where the value of $u$ is $65\%$ of the maximum value over the domain. Then we superpose these contours (as black solid lines) over the $u_{fil}$ field as shown in Fig.\,\ref{fig:5}($c$). We see that the black contours line up with the peaks in the patch of $u_{10}$ while valleys between them, indicated in blue, indicate the locations of missing $u_{01}$ peaks. It is also possible to discern distortions in the background $u_{01}$ state in the vicinity of the grain boundary. 
\begin{figure}
\centering{\includegraphics[width=0.45\linewidth]{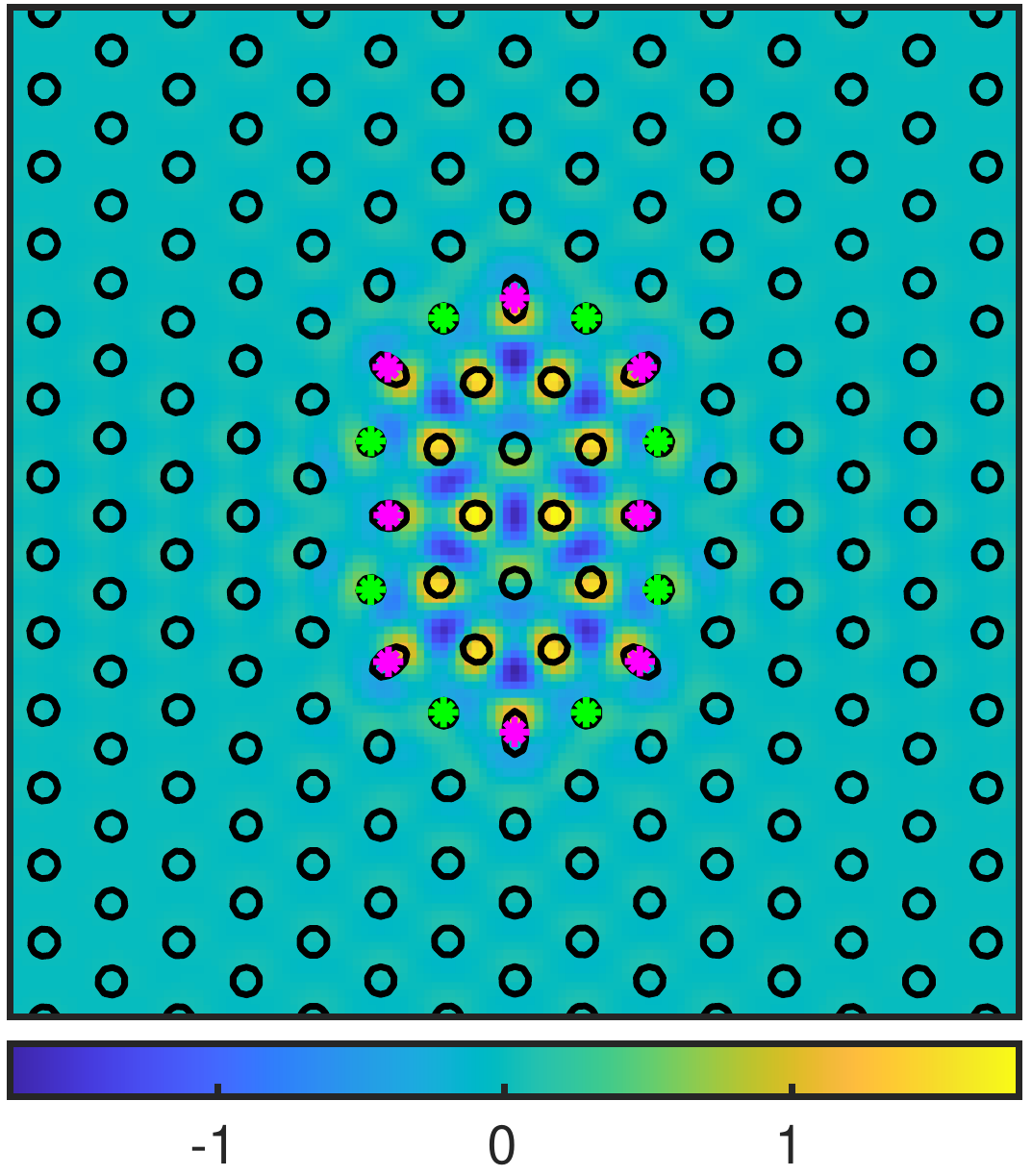} }
\caption{Contours of $u$ determined at $65\%$ of $\max(u)$ superposed on $u_{fil}$ (reproduced from Fig.\,\ref{fig:3}($c$). The locations of penta-hepta defects are indicated using green star markers for peaks with $5$ neighbours and pink star markers for peaks with $7$ neighbours. }
\label {fig:6}
\end{figure}

We seek to identify peaks that have five (seven) neighbours in order to locate penta-hepta defects automatically. In order to do this, we start with the data from the contours determined at $65\%$ of $\max(u)$ as seen in Fig.\,\ref{fig:5}($c$). We determine the grid $\mathcal{G}$ comprised of the centres of each closed black circle (representing a single peak) in the contour. For points in the grid $\mathcal{G}$, we create a 2D Delaunay triangulation and use it to determine the number of neighbours which surround each point in $\mathcal{G}$. From this, we can immediately identify the locations of peaks that have five (seven) neighbours. Figure \ref{fig:6} reproduces Fig.\,\ref{fig:5}($c$) with additional green star markers indicating peaks with five neighbours and pink star markers indicating peaks with seven neighbours. This same procedure is adopted to create figures comprising the contours of the field $u$ superposed on the image of $u_{fil}$ and overlaid with the locations of the penta-hepta defects in the movies associated with this paper. 


\section{To snake or not to snake}

Having obtained multiple states with penta-hepta defects as equilibria of the SH23 equation at a given parameter value (as shown in Fig.\,\ref{fig:3}), we seek to understand if these states are connected via homoclinic snaking, i.e., if they lie on the same solution branch. In order to determine this, we perform numerical continuation \citep{Doedel1991} of the solutions of the equation
\begin{equation}
\mu u - (1+\nabla^2)^2u + Qu^2 - u^3 = 0\,,
\label{eq:shnumcont}
\end{equation}
Here, as with time stepping, the field $u$ is discretised in space and nonlinear terms are calculated pseudo-spectrally. Pseudo-arclength continuation is performed with variable stepsize along the solution branch using a biconjugate gradient-stabilised method \citep{vandervorst1992} to approximate the action of the Jacobian of the system.  

\begin{figure}
\centering{\includegraphics[width=0.75\linewidth]{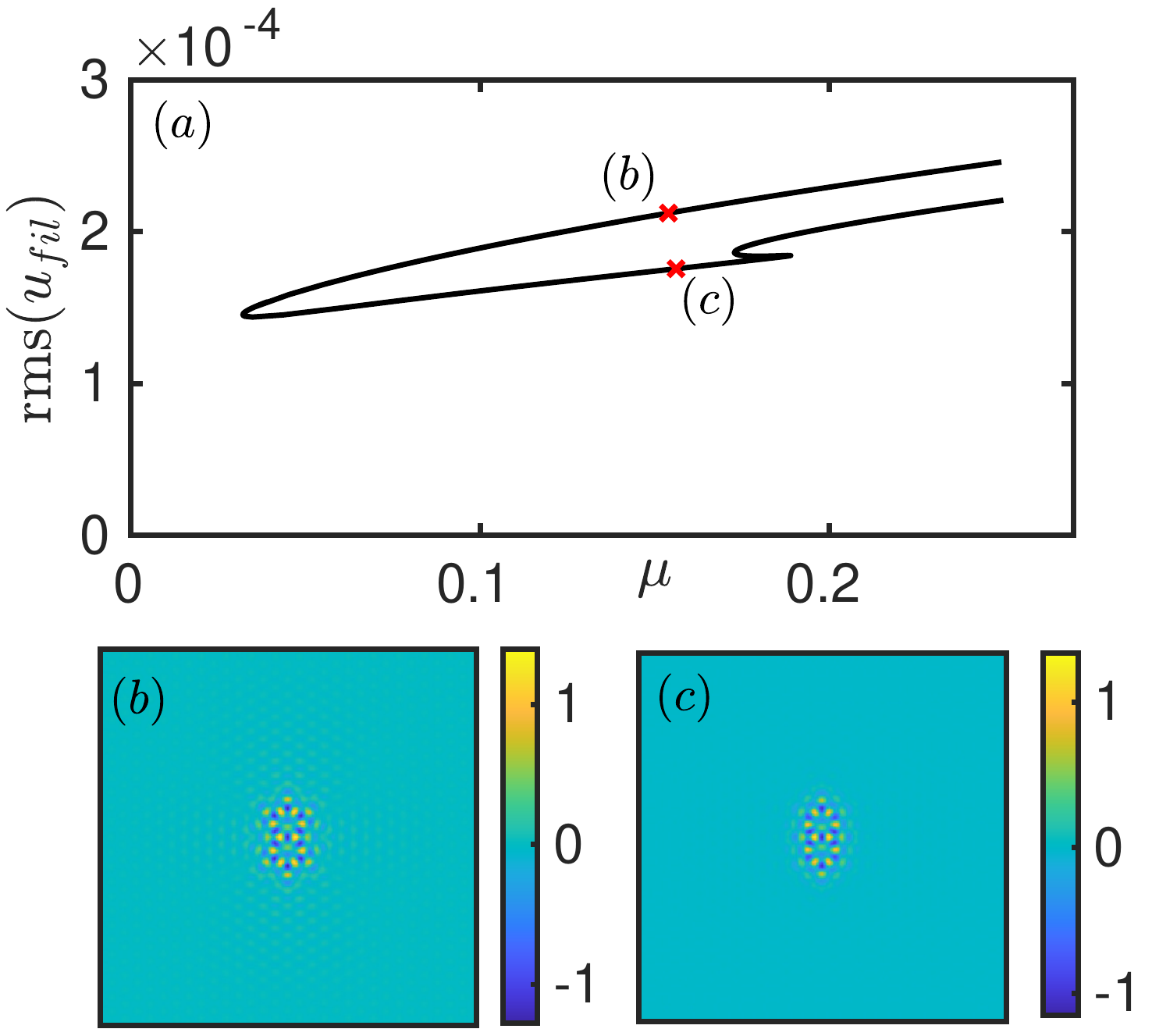} }
\caption{Snaking of states with penta-hepta defects shown in terms of ($a$) a bifurcation plot of rms($u_{fil}$) as a function of the bifurcation parameter $\mu$ and spectrally filtered fields $u_{fil}$ at points marked with a red cross in panel ($a$) with $\mu = 0.1539$ ($b$) and $\mu=0.1534$ ($c$), respectively. All other parameters as in Fig.\,\ref{fig:1}.}
\label {fig:7}
\end{figure}

We start with the state in Fig.\,\ref{fig:3}($a$) and vary the bifurcation parameter $\mu$. Figure \ref{fig:7} shows part of the solution branch starting from this state. With decreasing $\mu$, the measure $\textrm{rms}(u_{fil})$ decreases along the solution branch and the branch turns around at a saddle-node bifurcation at $\mu=0.032$, followed by two further turning points at $\mu=0.189$ and $\mu=0.173$ before returning to the starting value of $\mu=0.25$. In order to compare the states on the solution branch before and after the first saddle-node bifurcation, we pick two nearby values $\mu=0.1539$ and $\mu=0.1534$, labelled as ($b$) and ($c$) in Fig.\,\ref{fig:7}($a$) and display the corresponding fields $u_{fil}$ in Fig.\,\ref{fig:7}($b$) and ($c$), respectively. Between these two solutions, we observe that the localised patch of penta-hepta defects has lost four peaks and has become slightly smaller. Such a change, involving addition/removal of part of the pattern across a saddle-node bifurcation, is similar to changes observed during homoclinic snaking of localised hexagon states in a $u=0$ background in the subcritical regime $\mu<0$ \citep{LloydSIADS2008}. In addition the larger amplitude state ($b$) distorts the background state over a larger distance.  

\begin{figure}
\centering{\includegraphics[width=0.95\linewidth]{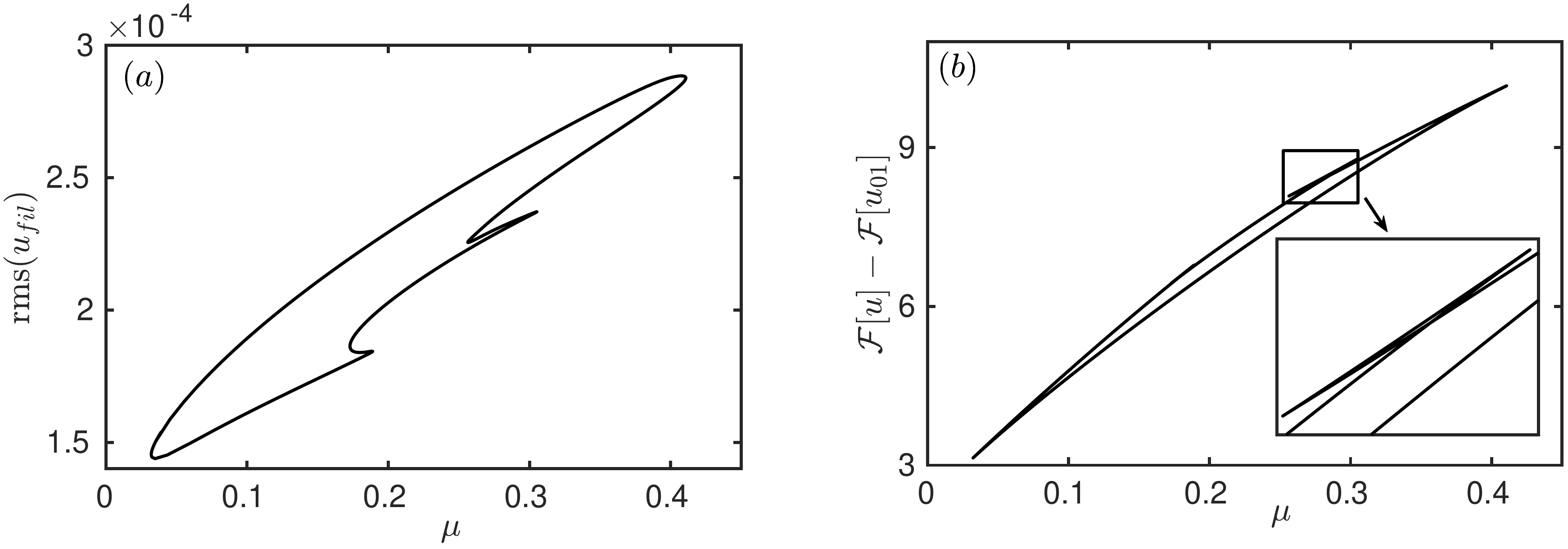} }
\caption{Complete isola of the solution branch related to the field in Fig.\,\ref{fig:5}($a$) shown in terms of ($a$) rms($u_{fil}$) and ($b$) the free energy difference, $\mathcal{F}[u] - \mathcal{F}[u_{01}]$, both as functions of the bifurcation parameter $\mu$. The inset in ($b$) shows an enlargement of the region within the rectangle displaying a swallow-tail behaviour in the solution branch.}
\label {fig:8}
\end{figure}

We track the full solution curve connected to the state in Fig.\,\ref{fig:3}($a$) and we find that for the chosen parameters we uncover an isola, which is displayed in Fig.\,\ref{fig:8}. Panel ($a$) shows the complete isola in terms of $\textrm{rms}(u_{fil})$ plotted as a function of the bifurcation parameter $\mu$ over the range $0.032<\mu<0.41$. Supplementary material Movie 1 shows how $\textrm{rms}(u_{fil})$ changes as we move over this isola. Note that the identification of peaks with five or seven neighbours is based on the 2D Delaunay triangulation, which in turn is based on the contour curves for the field $u$. This means that any asymmetry in the location of the associated penta-hepta defects we observe at multiple locations along the isola could be a consequence of two levels of discretisation: one at the level of determining the centre of a contour peak and the second at the level of the automated determination of the number of nearest neighbours from the resulting 2D Delaunay triangulation.

In the next panel, Fig.\,\ref{fig:8}($b$), we see another version of the same bifurcation diagram with a different norm on the $y$-axis. In order to understand the relative cost in terms of free energy that is needed to create the observed ring of penta-hepta defects, i.e., to calculate the grain boundary free energy \citep{hirvonen2016multiscale, hirvonen2017energetics}, we first determine the free energy, $\mathcal{F}[u_{01}]$, of the reference state $u_{01}$ from Eq.~\eqref{eq:free_energy}. We subtract this free energy from the free energy, $\mathcal{F}[u]$, corresponding to the field in Fig.\,\ref{fig:3}($a$). With this difference in free energy costs as our norm, the resulting isola is as shown in Fig.\,\ref{fig:8}($b$). As shown enlarged in the inset, this curve displays swallow-tail behaviour at locations where we observe two consecutive saddle-node bifurcations in Fig.\,\ref{fig:8}($a$). This result is similar to the occurrence of swallow-tail structures in the snaking of two-dimensional quasicrystals in \citet{SubramanianNJP2018}. The difference in free energy, i.e., $\mathcal{F}[u]-\mathcal{F}[u_{01}]$, gives the free energy of the interface and shows that this increases with $\mu$, since the perimeter of the grain hardly changes as we go around the isola. Of course, the grain boundary free energy depends on the relative orientation \citep{hirvonen2016multiscale} of the neighbouring grains, but this also does not change around the isola.
\begin{figure}
\centering{\includegraphics[width=0.95\linewidth]{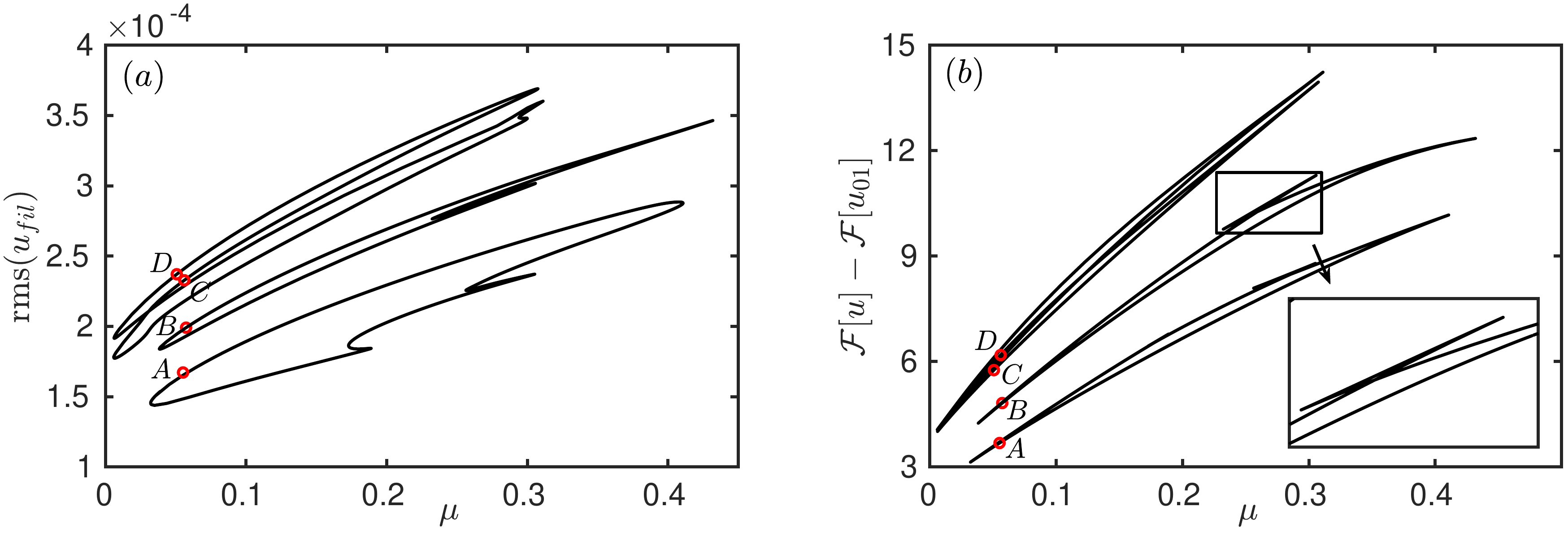} \\
\vspace{0.4cm}
\includegraphics[width=0.35\linewidth]{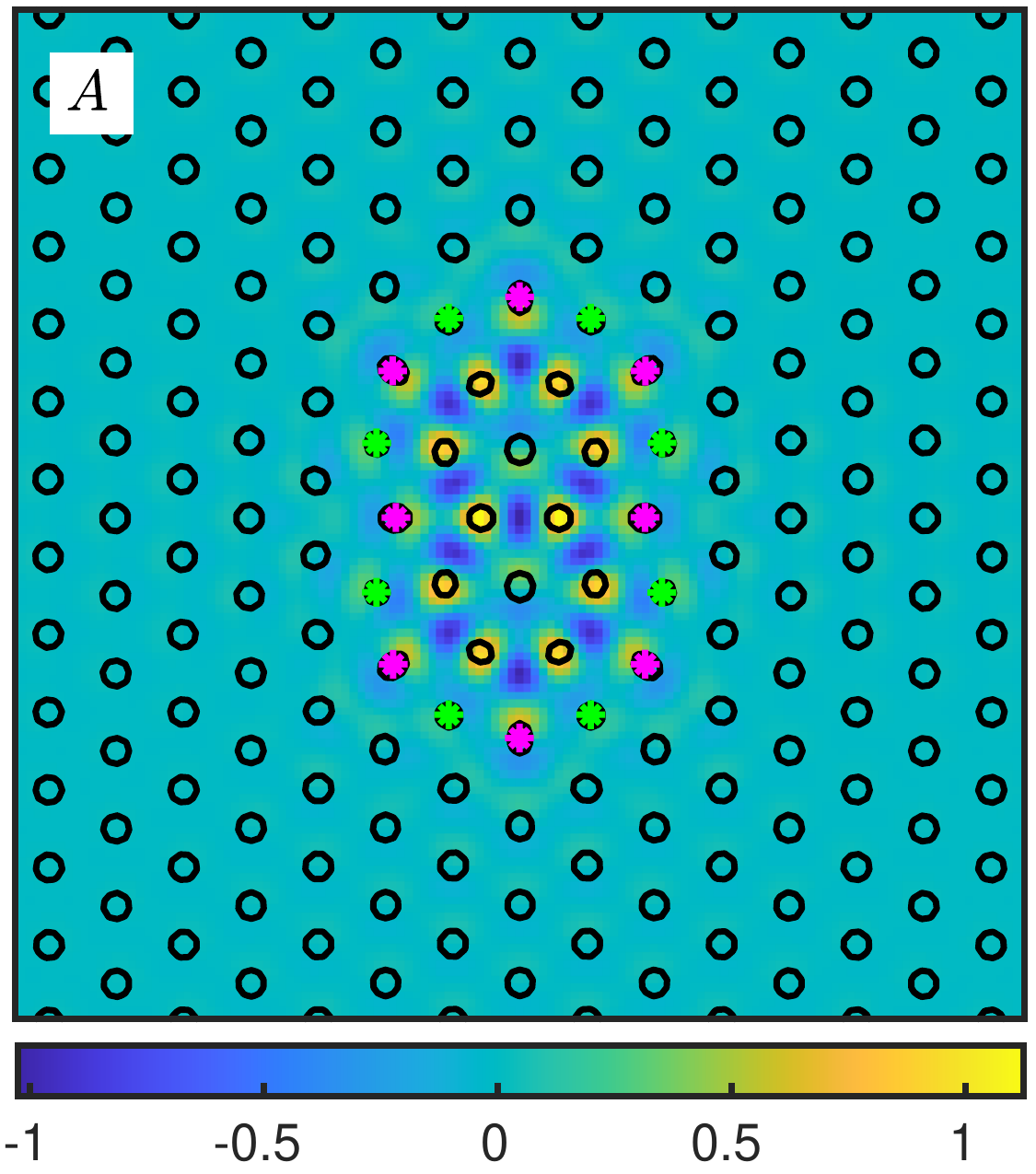} \hspace{0.5cm}
\includegraphics[width=0.35\linewidth]{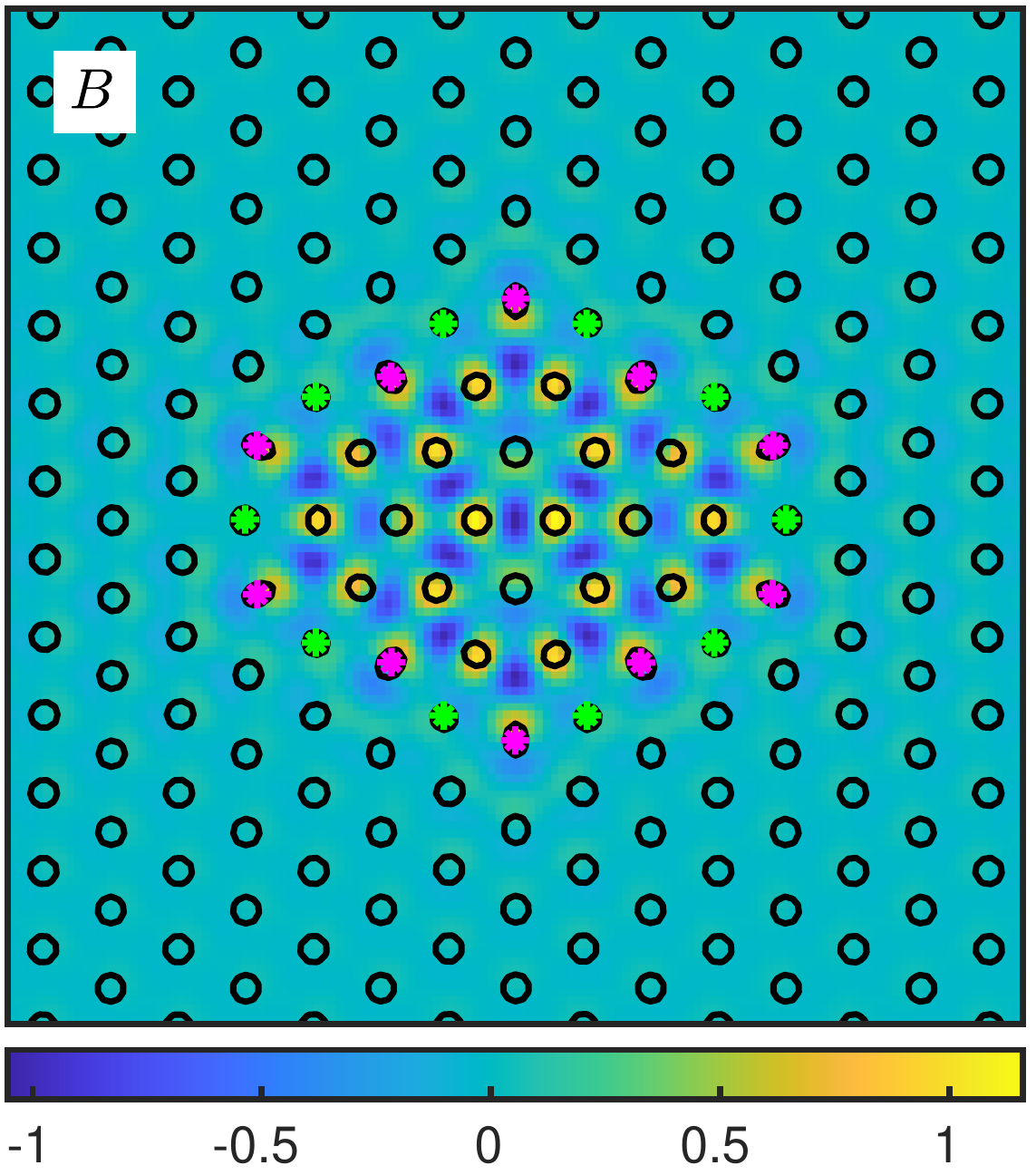} \\
\vspace{0.4cm}
\includegraphics[width=0.35\linewidth]{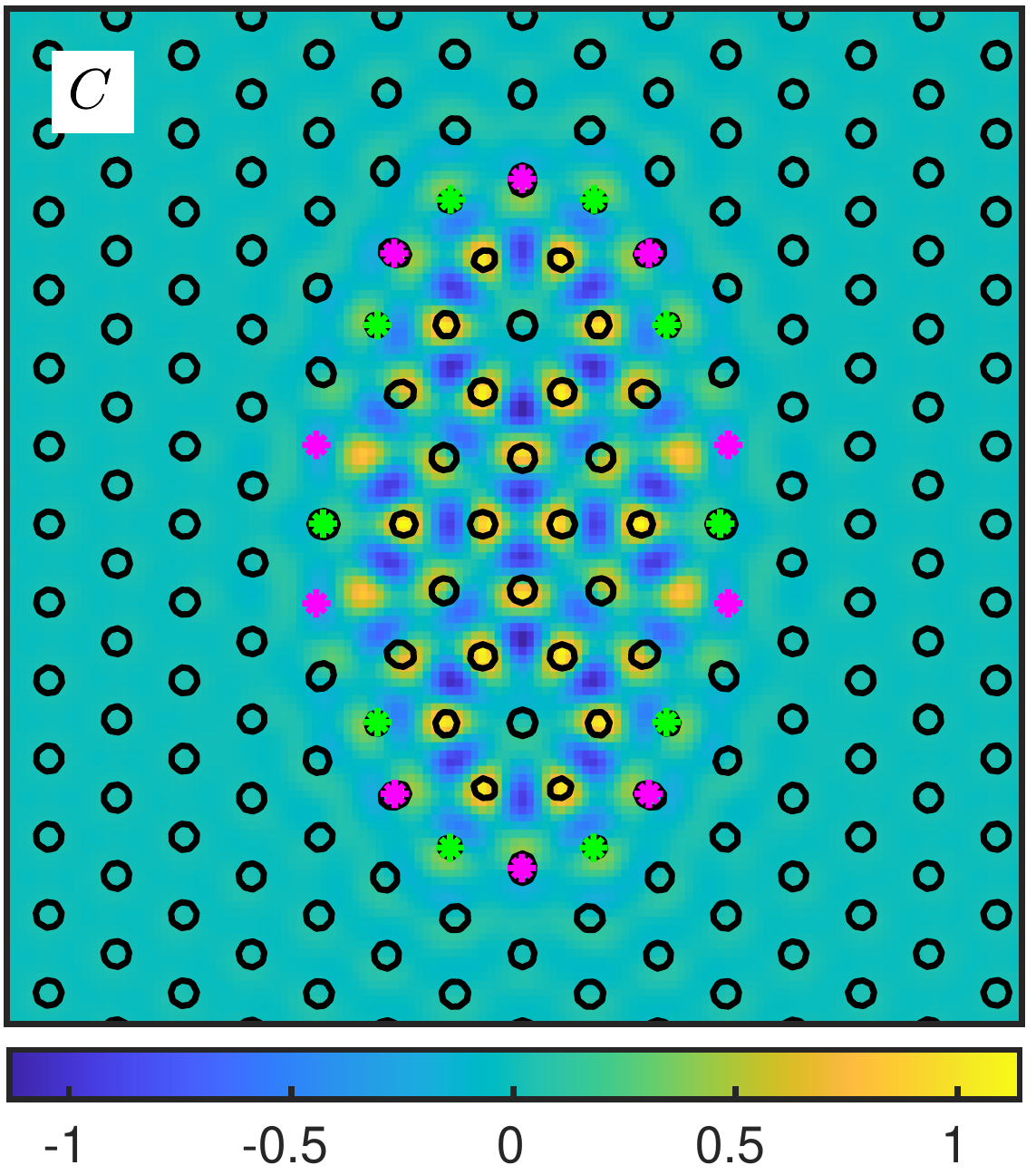} \hspace{0.5cm}
\includegraphics[width=0.35\linewidth]{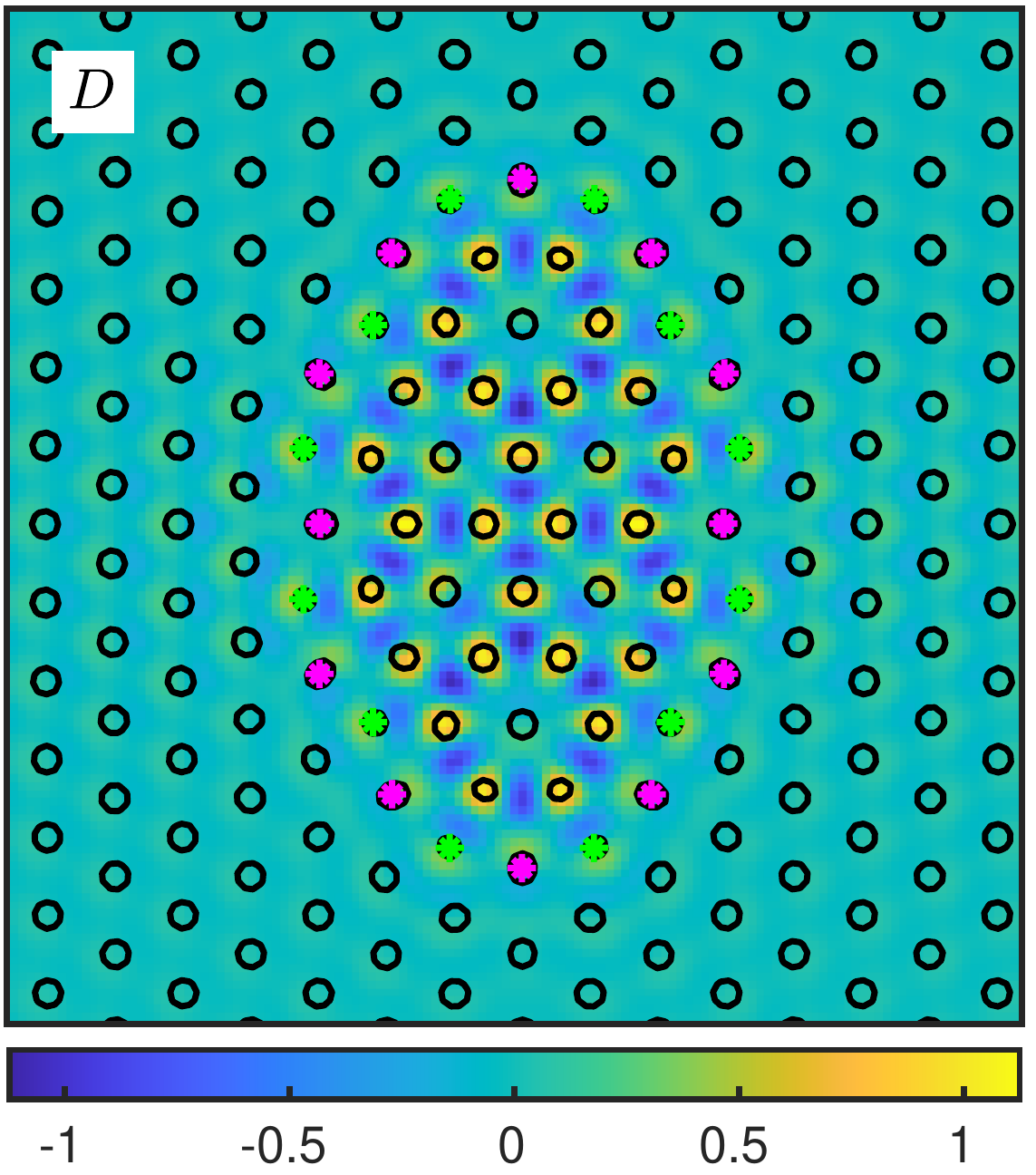}
}
\caption{Isolas of the solution branches related to the field in Fig.\,\ref{fig:5} shown in terms of ($a$) rms($u_{fil}$) and ($b$) the free energy difference, $\mathcal{F}[u] - \mathcal{F}[u_{01}]$, both as functions of the bifurcation parameter $\mu$. Panels $A$-$D$ show contours of the field $u$ superimposed with $u_{fil}$ state at points indicated with red circles in panels ($a$) and ($b$). The inset in panel ($b$) shows an enlargement of the black rectangle showing a swallow-tail in the isola connected to state $B$.}
\label {fig:9}
\end{figure}

Figure \ref{fig:9} shows the consolidated version of the previous figure with the solution branches associated with all the three states shown in Fig.\,\ref{fig:4} for $\mu=0.25$. Panel \ref{fig:9}($a$) shows the variation of rms($u_{fil}$) for the three isolas that result, all of which exist over a large range of the bifurcation parameter $\mu$. From this panel, we deduce a direct correlation between increased rms($u_{fil}$) values and the size of the central patch. Figure \ref{fig:9}($b$) shows the variation of the free energy difference between the extended hexagonal pattern and a state with a closed ring of PHDs, for the three different isolas. All of these free energy differences are positive, and increasing with the perimeter of the central patch. We observe the presence of multiple swallow-tail like structures along the isolas and we include an enlargement of one such swallow-tail from the isola connected to solution in Fig.\,\ref{fig:3}($b$). In this figure we also show additional solutions from each isola focusing on solutions at smaller values of $\mu$. The locations of each solution is marked with a red circle in panels Figs.\,\ref{fig:9}($a$-$b$) and identified with the letters $A$-$D$. Compared to Fig.\,\ref{fig:6}, we see that the variation of $u_{fil}$ in these panels is reduced, as reflected in the smaller difference between the maximum and minimum value of $u_{fil}$ in the associated colour bars. For a comprehensive look at how these solutions vary along different isolas, see the supplemental movie files. 

\begin{figure}
\centering{\includegraphics[width=0.9\linewidth]{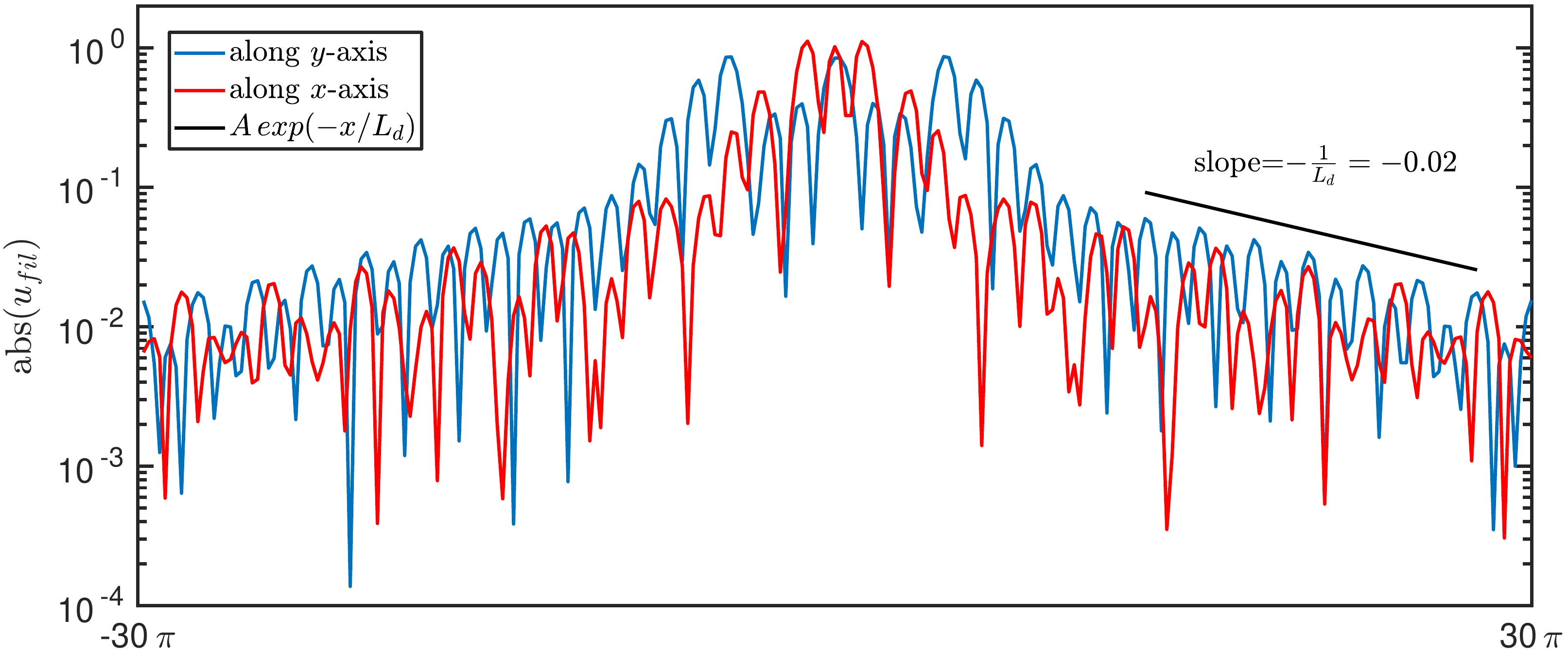} }
\caption{Transverse cuts of abs($u_{fil}$), i.e., the absolute value of the spectrally filtered field $u_{fil}$,  along two different directions, $x-$ and $y-$axis for the field $u$ from Fig.\,\ref{fig:9}, panel $A$. The straight line with slope $-1/L_{d}$ shows that outside the immediate grain, its influence falls off exponentially with a characteristic length scale $L_{d}\sim 50$.}
\label {fig:10}
\end{figure}

In order to understand the length scale over which the strain field generated by the defect decays, we look in Fig.\,\ref{fig:10} at the cross-section of the filtered field $u_{fil}$ along two different directions and do so for the field shown in Fig.\,\ref{fig:9}, panel $A$. Cuts of abs($u_{fil}$), i.e., the absolute value of the spectrally filtered field $u_{fil}$, along the $x-$ and $y-$directions both show large values at the centre of the domain with long oscillating exponential tails. From the slope of the decay we estimate the decay length $L_{d}$ to be 50, i.e., several times the pattern length scale ($2\pi$). We mention that exponential localisation is expected of the subcritical regime $\mu<0$, even in sheared situations (\citet{Brand1}; \citet{Brand2}) but not necessarily for the localised axisymmetric spots that are present in SH35 even in the supercritical regime $\mu>0$ \citep{LloydSNon2009}. 

\begin{figure}
\centering{\includegraphics[width=0.85\linewidth]{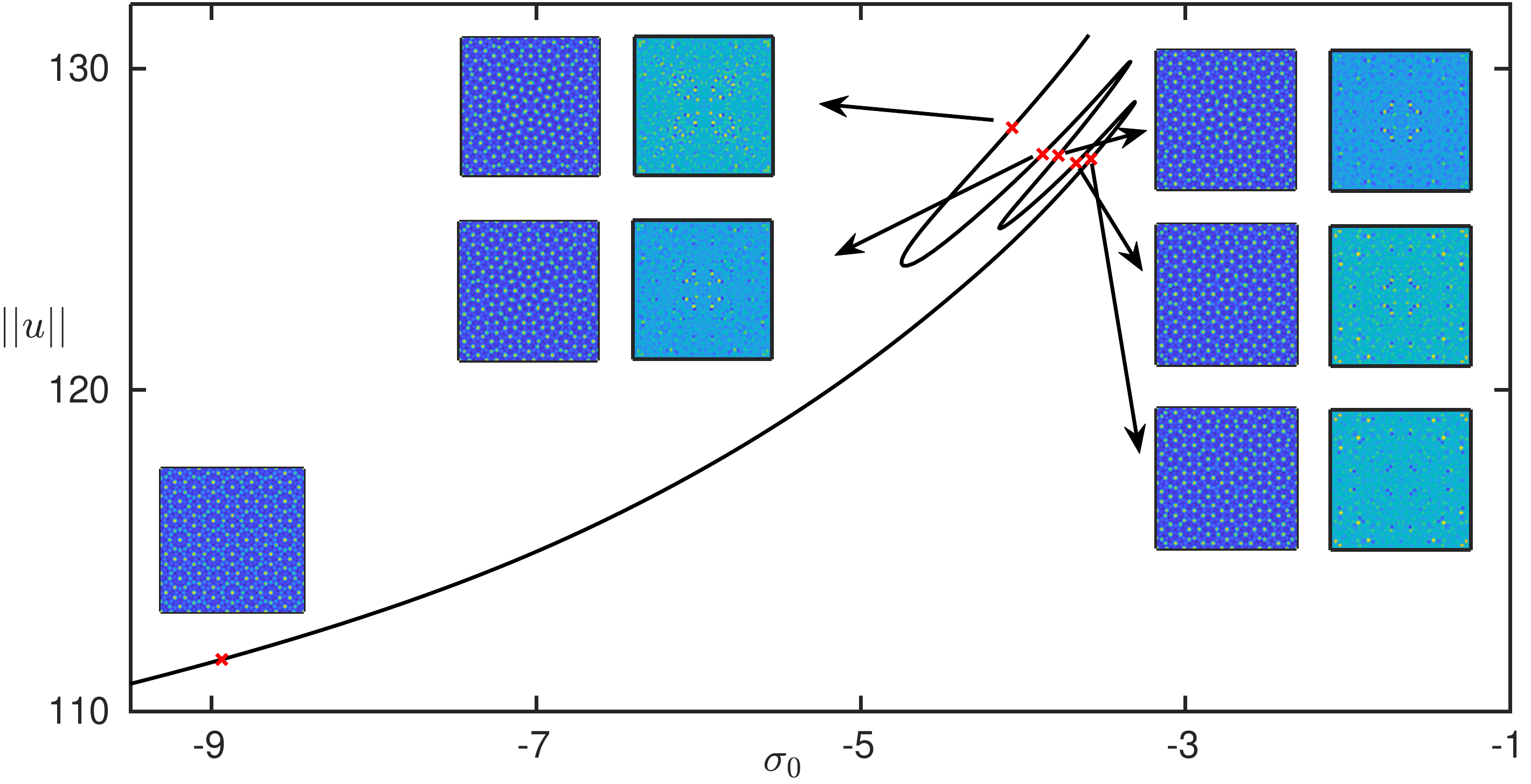} }
\caption{Solution branch of dodecagonal quasicrystals in a conserved PFC model for soft matter crystallisation showing the norm of the density-like field $u$ as a function of the bifurcation parameter $\sigma_0$. Inset at bottom left corner is the chosen reference state with $\sigma_0=-8.94$. Insets indicated by arrows are the full field (left) and the filtered field $u_{fil}$ relative to the reference state (right) at points marked with red crosses along the snaking solution branch. Values of the other parameters in the model are: $\mu=\nu=-0.412$, $Q=2$ and $q=(1/2)\cos(\pi/12)$. For details of the model, see \citet{Subramanian2016}.}
\label{fig:11}
\end{figure}

We next turn to an example where defect states do snake, instead of lying on a stack of isolas, as in Fig.~\ref{fig:9}. Our example comes from a related but mass-conserving phase field crystal (PFC) model that forms patterns at two length scales. The model describes crystallisation of soft matter into complex patterns and is closely related to the Swift--Hohenberg equation \eqref{eq:sh23} suitably modified to generate instability at two length scales instead of just one. The model has been shown to produce both extended and spatially localised dodecagonal quasicrystals in 2D and icosahedral quasicrystals in 3D. See \citet{Subramanian2016} and \citet{SubramanianNJP2018} for details. The model describes the evolution of a scalar density-like field $u({\bf x})$ and contains a parameter $\sigma_0$ that controls length scale selectivity, with more negative values of $\sigma_0$ corresponding to sharper length scale selectivity. As we follow the branch of fully nonlinear dodecagonal quasicrystal solutions (Fig.~\ref{fig:11}, leftmost inset) for increasing $\sigma_0$ the system starts to form localised hexagonal inclusions within the quasicrystal representing defects. Once such defects are present, the solution branch undergoes snaking  as shown in Fig.\,\ref{fig:11} where the norm $||u||$ is plotted as a function of $\sigma_0$.

The RLVs for a quasicrystalline pattern are drawn from two circles of wavevectors and the higher order RLVs introduce new length scales at both ever larger wavevectors and at wavevectors close to zero, including wavevectors close to the first order RLVs \citep{rucklidge2003,SNLT2020}. Therefore, the construction of the correct filter matrix $\mathcal{P}$ in the case of a quasicrystalline reference pattern is rather more complex. We therefore choose a fixed reference state at $\sigma_0=-8.94$ and use this state to visualise the changes that occur in the field $u$ with increasing $\sigma_0$. We then identify points along different parts of the snaking curve (identified by red crosses in Fig.\,\ref{fig:11}) where we plot the field $u$ (left panels) side-by-side with the spectrally filtered rms($u_{fil}$) calculated with respect to the reference state. 

The left inset panels in each pair show that as we move higher up the snaking branch more of the region is covered by hexagonal patches, while the right inset panels display prominent peaks at locations where there is a change from quaiscrystalline to hexagonal local ordering. We note that these right panels are not as flat as the rms($u_{fil}$) states that we obtained  in the SH23 system for PHDs. This is due to several reasons: the complex way in which RLVs add new length scales for quasicrystalline patterns at higher orders, the choice of the reference state, etc. As a result this analysis remains incomplete but we use it here to highlight the fact that fully nonlinear equilibria with defects can indeed snake, forming multiple coexisting defect structures over a parameter interval, and leave a more detailed study of this system to future work.


\section{Conclusion and future work}

In this paper we have demonstrated by explicit computation on a model problem, the SH23 equation in the plane, that defects in a background hexagonal pattern come in discrete families, as expected of systems in which pinning plays an important role. In particular, we have shown that these defects lie on closed solution curves called isolas and so do not snake, in contrast to the behaviour of localised states in the subcritical parameter regime in SH23 \citep{burke2006localized,LloydSIADS2008}. In other cases, such the two-scale PFC model, the defects are associated with the presence of classical snaking behaviour.

These results suggest that defects in patterns, whether periodic or quasiperiodic, behave much like localised structures on a spatially uniform background and that they can be studied using the same techniques from pattern formation theory as already used very effectively in the study of other states. Careful demodulation, that is, removal of the background pattern, appears to be key to identifying defects in simulations or experiments, although the nature of the background state brings its own properties into the behaviour of these states: pinning to background microstructure. This effect must be retained in any theoretical description of these defect states, precluding the use of the commonly invoked amplitude-phase decomposition.

The defects studied here are associated with grain boundaries and we have introduced these into the system by rotating the hexagonal pattern in a small part of the domain through 30$^{\circ}$. This angle determines the number of PHDs associated with the grain boundary, and hence the structure of this boundary \citep{hirvonen2016multiscale}. Since this structure is associated with the inclusion of one pattern within another, both of which have the same free energy when separately covering the plane, there is no bulk phase free energy difference to drive the growth or shrinkage of the grain boundary between the two regions of different orientation. Instead, all the energy difference between the state with two grains and a defect-free pattern is associated with the boundary itself, and in particular the PHDs introduced into the pattern. It is noteworthy that this energy is positive (Fig.~\ref{fig:8}($b$)) indicating that such defect structures will not form spontaneously but that they require finite amplitude perturbations to initiate them.

The fact that the grain boundaries are associated with a positive free energy implies that the system will seek to eliminate such boundaries as much as possible -- this is, by grain rotation. Since many of the observed states discussed above are dynamically stable, this implies that for the parameter values where these isolas exist, there is a free energy barrier that the system must surmount to decrease the inner grain size. In the materials science context these results show that in this parameter regime grain shrinkage is a thermally activated process. Note that PFC models have previously been applied to study grain rotation \citep{radhakrishnan2012comparison, huter2017modelling} and also other, more complex grain boundary and defect dynamics that occur, for example, when a material is sheared \citep{chan2010plasticity}. Our results identify the parameter regime where grains in a polycrystalline material are dynamically stable and so identifies the regime in which grain rotation is an activated process. In addition, the different states identified by our approach will be attractors in the dynamics of a polycrystalline material equilibrating over time \citep{trautt2012grain}.

Related to this, we emphasise a point that can clearly be seen from in Fig.~\ref{fig:8}: the localised grain defect structures persist over a large range of values of the bifurcation parameter $\mu$. We ascribe this fact to the absence of a unique Maxwell point between the two competing patterns (recall that both have the same free energy), thereby reducing the tendency towards depinning. In related systems possessing a Maxwell point, such as SH23 in the subcritical regime, the pinning region is limited by the onset of depinning that leads to the time-dependent invasion of the higher energy state by the lower energy state. We interpret Fig.~\ref{fig:8} as suggesting that the presence of adjacent PHDs, and in particular the observed ring structure, stabilises individual PHDs and suggest that this fact may explain our inability to generate isolated PHDs in this system.

Our results also reveal that the localised defect structures influence the background pattern over a region larger than the defect itself. For example Fig.\,\ref{fig:7}($b$) clearly shows that the perturbations in $u_{fil}$ persist over a large part of the domain. This indicates that the perturbations due to the presence of the defects decay over length scales that are several times larger than the typical distance between peaks in the pattern. Figure \ref{fig:10} allows us to see precisely the decay length $L_{d}$ of these correlations, showing that they falls off exponentially with distance from the defect structure.

This paper represents a first step towards understanding the multiplicity of defect states in two spatial dimensions. A number of questions remain, including
\begin{itemize}
\item What is the smallest stationary and/or dynamically stable defect structure that can exist?
\item Can we develop other topological measures to determine local order in a pattern with defect \citep{MOTTA201817}.
\item How do the defect structures depend on the angle between the two hexagonal patterns \citep{hirvonen2017energetics} and how does this angle affect the snaking properties of these states? For example, is it possible to uncover angles for which the defect states snake, rather than lying on disjoint isolas?
\item What happens if non-variational terms are incorporated in the model equation? Does this generate motion of the defect state or do the defects remain stationary and stable?
\item More generally, what are implications of these structures for conserved systems and materials science in general?
\end{itemize}
We leave these questions for future work on this rich and fruitful class of problems. 


\section*{Acknowledgment}
The authors would like to acknowledge interesting discussions with David Lloyd, Herman Riecke and Yuan-Nan Young. This work was supported in part by a Hooke Research Fellowship (PS), the National Science Foundation under grant DMS-1908891 (PS, EK) and the Engineering and Physical Sciences Council under the grants EP/P015689/1 (AJA) and EP/P015611/1 (AMR).  

%
%

\bibliographystyle{imamat}
\bibliography{sample}

\end{document}